\documentstyle{mn}

\newif\ifAMStwofonts
\def \SAIT #1 #2 {{\em Mem.\ Soc.\ Astron.\ It.\/} {\bf #1}, #2}
\def \MESS #1 #2 {{\em The Messenger\/} {\bf #1}, #2}
\def \ASTRNACH #1 #2 {{\em Astron. Nach.\/} {\bf #1}, #2}
\def \AAP #1 #2 {{\em Astron. Astrophys.\/} {\bf #1}, #2}
\def \AAL #1 #2 {{\em Astron. Astrophys. Lett.\/} {\bf #1}, L#2}
\def \AAR #1 #2 {{\em Astron. Astrophys. Rev.\/} {\bf #1}, #2}
\def \AAS #1 #2 {{\em Astron. Astrophys. Suppl. Ser.\/} {\bf #1}, #2}
\def \AJ #1 #2 {{\em Astron. J.\/} {\bf #1}, #2}
\def \ANNREV #1 #2 {{\em Ann. Rev. Astron. Astrophys.\/} {\bf #1}, #2}
\def \APJ #1 #2 {{\em Astrophys. J.\/} {\bf #1}, #2}
\def \APJL #1 #2 {{\em Astrophys. J. Lett.\/} {\bf #1}, L#2}
\def \APJS #1 #2 {{\em Astrophys. J. Suppl.\/} {\bf #1}, #2}
\def \APSS #1 #2 {{\em Astrophys. Space Sci.\/} {\bf #1}, #2}
\def \ASR #1 #2 {{\em Adv. Space Res.\/} {\bf #1}, #2}
\def \BAIC #1 #2 {{\em Bull. Astron. Inst. Czechosl.\/} {\bf #1}, #2}
\def \JSQRT #1 #2 {{\em J. Quant. Spectrosc. Radiat. Transfer\/} {\bf #1}, #2}
\def \MN #1 #2 {{\em Mon. Not. R. Astr. Soc.\/} {\bf #1}, #2}
\def \MEM #1 #2 {{\em Mem. R. Astr. Soc.\/} {\bf #1}, #2}
\def \PLR #1 #2 {{\em Phys. Lett. Rev.\/} {\bf #1}, #2}
\def \PASJ #1 #2 {{\em Publ. Astron. Soc. Japan\/} {\bf #1}, #2}
\def \PASP #1 #2 {{\em Publ. Astr. Soc. Pacific\/} {\bf #1}, #2}
\def \NAT #1 #2 {{\em Nature\/} {\bf #1}, #2}
\def\strutdepth{\dp\strutbox}
\def\marginalstar{\strut\vadjust{\kern-\strutdepth\specialstar}}
\def\specialstar{\vtop to \strutdepth{\baselineskip\strutdepth\vss\llap{$\bigtriangleup\mskip-12mu\bigtriangledown$ }\null}}
\def \nm {N_{\mathrm{m}}}
\def \nm0 {N_{\mathrm{m,0}}}
\def \na {N_{\mathrm{a}}}
\def \na0 {N_{\mathrm{a,0}}}
\def \sm {\sigma_{\mathrm{m}}}

\def \BGE {\begin{equation}}
\def \EDE {\end{equation}}

\ifoldfss
  \ifCUPmtlplainloaded \else
    \NewTextAlphabet{textbfit} {cmbxti10} {}
    \NewTextAlphabet{textbfss} {cmssbx10} {}
    \NewMathAlphabet{mathbfit} {cmbxti10} {} % for math mode
    \NewMathAlphabet{mathbfss} {cmssbx10} {} %  "   "    "
  \fi
  \ifAMStwofonts
    \ifCUPmtlplainloaded \else
      \NewSymbolFont{upmath} {eurm10}
      \NewSymbolFont{AMSa} {msam10}
      \NewMathSymbol{\upi}     {0}{upmath}{19}
      \NewMathSymbol{\umu}     {0}{upmath}{16}
      \NewMathSymbol{\upartial}{0}{upmath}{40}
      \NewMathSymbol{\leqslant}{3}{AMSa}{36}
      \NewMathSymbol{\geqslant}{3}{AMSa}{3E}

      \let\leq=\leqslant \let\le=\leqslant
       
    \fi
  \fi
\fi % End of OFSS

\ifnfssone
  \newmathalphabet{\mathit}
  \addtoversion{normal}{\mathit}{cmr}{m}{it}
  \addtoversion{bold}{\mathit}{cmr}{bx}{it}
  \newmathalphabet{\mathbfit} % math mode version of \textbfit{..}
  \addtoversion{normal}{\mathbfit}{cmr}{bx}{it}
  \addtoversion{bold}{\mathbfit}{cmr}{bx}{it}
  \newmathalphabet{\mathbfss} % math mode version of \textbfss{..}
  \addtoversion{normal}{\mathbfss}{cmss}{bx}{n}
  \addtoversion{bold}{\mathbfss}{cmss}{bx}{n}
  \ifAMStwofonts
    \ifCUPmtlplainloaded \else
      \UseAMStwoboldmath
      \makeatletter
      \new@mathgroup\upmath@group
      \define@mathgroup\mv@normal\upmath@group{eur}{m}{n}
      \define@mathgroup\mv@bold\upmath@group{eur}{b}{n}
      \edef\UPM{\hexnumber\upmath@group}
      \new@mathgroup\amsa@group
      \define@mathgroup\mv@normal\amsa@group{msa}{m}{n}
      \define@mathgroup\mv@bold\amsa@group{msa}{m}{n}
      \edef\AMSa{\hexnumber\amsa@group}
      \makeatother
      \mathchardef\upi="0\UPM19
      \mathchardef\umu="0\UPM16
      \mathchardef\upartial="0\UPM40
      \mathchardef\leqslant="3\AMSa36
      \mathchardef\geqslant="3\AMSa3E

      \let\leq=\leqslant \let\le=\leqslant

    \fi
  \fi
\fi % End of NFSS release 1

\ifnfsstwo
  \DeclareMathAlphabet{\mathbfit}{OT1}{cmr}{bx}{it}
  \SetMathAlphabet\mathbfit{bold}{OT1}{cmr}{bx}{it}
  \DeclareMathAlphabet{\mathbfss}{OT1}{cmss}{bx}{n}
  \SetMathAlphabet\mathbfss{bold}{OT1}{cmss}{bx}{n}
  \ifAMStwofonts
    \ifCUPmtlplainloaded \else
      \DeclareSymbolFont{UPM}{U}{eur}{m}{n}
      \SetSymbolFont{UPM}{bold}{U}{eur}{b}{n}
      \DeclareSymbolFont{AMSa}{U}{msa}{m}{n}
      \DeclareMathSymbol{\upi}{0}{UPM}{"19}
      \DeclareMathSymbol{\umu}{0}{UPM}{"16}
      \DeclareMathSymbol{\upartial}{0}{UPM}{"40}
      \DeclareMathSymbol{\leqslant}{3}{AMSa}{"36}
      \DeclareMathSymbol{\geqslant}{3}{AMSa}{"3E}

      \let\leq=\leqslant \let\le=\leqslant

    \fi
  \fi
\fi % End of NFSS release 2

\ifCUPmtlplainloaded \else
  \ifAMStwofonts \else % If no AMS fonts
    \def\upi{\pi}
    \def\umu{\mu}
    \def\upartial{\partial}
  \fi
\fi
   \title[The artificial night sky brightness mapped from DMSP-OLS measurements]{The artificial night sky brightness mapped from DMSP satellite Operational Linescan System measurements}

   \author[P. Cinzano,        
          F. Falchi,
          C. D. Elvidge and 
          K. E. Baugh        
         ]{P. Cinzano$^{1}$\thanks{E-mail: cinzano@pd.astro.it},        
          F. Falchi$^{1}$,
          C. D. Elvidge$^2$ and
          K. E. Baugh$^2$\\
$^{1}$ Dipartimento di Astronomia, Universit\`a di Padova,
vicolo dell'Osservatorio 5,  I-35122 Padova, Italy\\
$^2$ Solar-Terrestrial Physics Division, NOAA National Geophysical Data Center, 3100 Marine Street, Boulder CO 80303}

\date{Accepted 19xx December xx.
      Received 19xx December xx;
      in original form 19xx October xx}

\pagerange{\pageref{firstpage}--\pageref{lastpage}}
\pubyear{1994}

\input epsf.sty
\begin{document}

\maketitle

\label{firstpage}

 \begin{abstract} 

We present a method to map the artificial sky brightness across large territories in astronomical photometric bands with a resolution of approximately 1 km. This is useful to quantify the situation of night sky pollution, to recognize potential astronomical sites and to allow future monitoring of trends.
The artificial sky brightness present in the chosen direction at a given position on the Earth's surface is obtained by the integration of the contributions produced by every surface area in the surrounding. Each contribution is computed  based on detailed models for the propagation in the atmosphere of the upward light flux emitted by the area.
The light flux is measured with top of atmosphere radiometric observations made by the Defense Meteorological Satellite Program (DMSP) Operational Linescan System. 

We applied the described method to Europe obtaining the maps of artificial sky brightness in V and B bands.

\end{abstract}

     
%
%________________________________________________________________

\section{Introduction}

The night sky is a world heritage.  In recent decades there has been a rapid increase in the brightness of the night sky in nearly all countries.  This has degraded astronomical viewing conditions.  The increase in night sky brightness is one of the most noticeable effects  of {\em light pollution}, which can be defined as the alteration of natural light levels in the outdoor environment due to artificial light sources.  The widespread use of artificial lighting with little regard to fixture shielding or energy
conservation is the primary source of anthropogenic light contributing to the brightness of the night sky. The astronomical community has expressed its concern over the growth of the sky brightness in a number of official documents and positions (e.g. the Resolutions of the General Assemblies of the International Astronomical Union (IAU)  XVI/9 1976, XIX/B6 1985, XX/A2 1988, XXIII/A1 1997 and the Positions of the American Astronomical Society). The UNESCO, the United Nations and the Commission Internationale de l'Eclairage (CIE) give consideration to this concern. The Commission 50 of IAU (The protection of existing and potential astronomical sites) is working actively in order to preserve the astronomical sky and many studies and meetings have been dedicated to this topic (e.g. Crawford 1991; Kovalevsky 1992; McNally 1994;  Isobe \& Hirayama 1998; Cinzano 2000a; see also Cinzano 1994 for a large reference list).
Laws, bills, standard rules, ordinances, regulations limit in many countries  the direct wasting of light in the sky from lighting fixtures and, in some cases, also the quantity of light reflected by lighted surfaces. The International Dark-Sky Association is active worldwide in this battle of culture and intelligence with the aim of building awareness and saving for mankind the possibility of feeling part of the universe. 

An effective battle against light pollution requires the knowledge of the condition of the night sky across large territories, recognition of vulnerable areas, the determination of the growth trends and the identification of most polluting cities. Therefore  a method to map the artificial sky brightness across large territories is required. This is also useful in order to recognize areas with low level of light pollution and potential astronomical sites.
In the past, mappings of sky brightness for extended areas has been performed based on population density data with some simple modelling. Examples include the works of Walker (1970, 1973) in California and Arizona, Albers \cite{alb} in USA, Bertiau et al. \cite{bert} in Italy, Berry \cite{berry} and Pike \& Berry \cite{pike} in Ontario (Canada). These authors used population data of cities to estimate their upward light emission and a variety of empirical or semi-empirical propagation laws for light pollution in order to compute the sky brightness produced by them. 
Recently advances in the availability and gain control of the DMSP satellite now provides direct measurements of the  upward light emission from nearly the entire surface of the Earth (60\degr South to 72\degr North). The first global DMSP image of the Earth at night was produced at 10 km resolution  (Sullivan 1989, 1991) using DMSP film strips, the only data available at that time.  Beginning in 1992, the availability of digital DMSP data has enabled a new set of higher spatial resolution products (Elvidge et al.\ 1997a, 1997b, 1997c).  These data have been used to model population distribution (Sutton et al.\ 1997) and urban sprawl impacts on food production (Imhoff et al.\ 1997a, 1997b).  The data have also been used to document light pollution as expressed in the increase in the upward flux of light over time for Japan (Isobe \& Hamamura 1998).  

In this paper we present a method to map the artificial sky brightness across large territories.  In order to bypass errors  arising when using population data to estimate upward flux, we construct the maps based on direct measuraments of the upward flux as observed from space using DMSP satellite night-time images and compute the downward flux to the Earth's surface with detailed modelling of  light pollution propagation in the atmosphere. We also present, as an application, detailed maps of artificial sky brightness in Europe in V and B astronomical photometrical bands with a resolution of approximately 1 km.
 In section \ref{ss} we describe the OLS-DMSP satellite observations, their reduction and analysis.  We also discuss
 the relation upward flux versus city population.
 In section \ref{s4} we describe the mapping technique and in section \ref{s5} we discuss the application to the maps of Europe. The maps are presented in section \ref{s6} together with our comments. Section \ref{s7} contains our conclusions.

\section{Observations and data analysis}
\label{ss}
\subsection{Satellite data}
\label{s1}

U.S. Air Force Defense Meteorological Satellite Program (DMSP) satellites are in low altitude (830 km) sun-synchronous polar orbits with an orbital period of 101 minutes. With 14 orbits per day they generate a global night time and day time coverage of the Earth every 24 hours with their main purpose to monitor the distribution of clouds and to assess navigation conditions. 
The U.S. Department of Commerce, NOAA National Geophysical Data Center (NGDC) serves as the archive for the DMSP program.  The digital archive was initiated in 1992.  Starting in 1994, NGDC has embarked on the development of night time lights processing algorithms and products from the DMSP-OLS (Elvidge et al.\ 1997a, 1997b, 1997c).

\begin{figure}
\epsfysize=7.8cm % fix the y-dimension and scales x-dim. to y-dim.
\hspace{0cm}\epsfbox{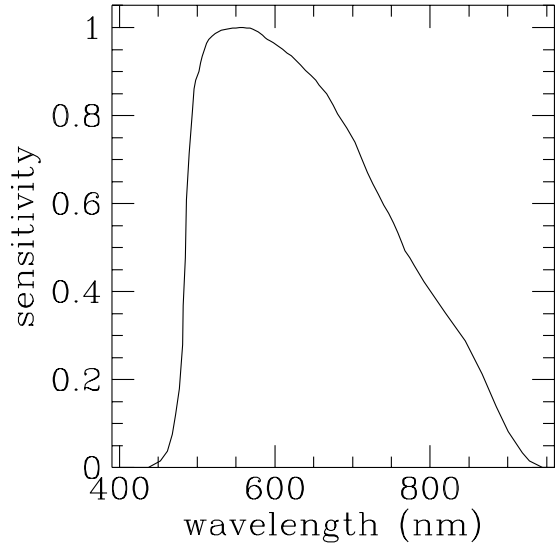} %for centering: act on hspace argument 
\caption[h]{Spectral sensitivity of OLS-PMT detector of DMSP satellite F12 .}
\label{fig1}
\end{figure}

The Operational Linescan System (OLS) is an oscillating scan radiometer with low-light visible and thermal infrared (TIR) imaging capabilities which first flew on DMSP satellites in 1976. 
At night the OLS uses a Photo Multiplier Tube (PMT) to intensify the visible band signal.  The purpose of this intensification is to observe clouds illuminated by moonlight. The PMT data have a broad spectral response from 440 to 940 nm (485 - 765 nm FWHM) with highest sensitivity in the 500 to 650 nm region (see Fig. \ref{fig1}). This covers the range for primary emissions from the most widely used lamps for external lighting:  Mercury Vapour (545 nm and 575 nm), High Pressure Sodium (from 540 nm to 630 nm), Low Pressure Sodium (589 nm). 
The sensitivity of the PMT, combined with the OLS-VDGA (variable digital gain amplifier) and fixed gain OLS pre- and post- amplifiers allows measurement of radiances down to  $10^{-10}$ W cm$^{-2}$ sr$^{-1}$ $\mu$m$^{-1}$ (Elvidge et al. 1999). This implies that the OLS-PMT could detect radiation with an effective wavelength near 550 nm down to a luminance of approximately  0.2 mcd m$^{-2}$. The TIR detector is sensitive  to radiation from 10.0 to 13.4 $\mu$m (10.3 - 12.9 FWHM).

The OLS acquires swaths of data that are 3000 km wide.  The OLS sinusoidal scan maintains a nearly constant along-track pixel-to-pixel ground sample distance (GSD) of 0.56 km.  Likewise, the electronic sampling of the signal from the individual scan lines maintains a GSD of 0.56 km.  The effective instantaneous field of view (EIFOV) is larger than the 0.56 km GSD and varies with scan angle.  At nadir the EIFOV is 2.2 km and expands to 5.4 km at the edge of scan 1500 km out from nadir.  Most of the data received by NGDC has been "smoothed" by on-board averaging of 5 $\times$ 5 pixel blocks, yielding data with a GSD of 2.8 km.  Other details of the OLS are described  by Lieske \cite{lieske}. 

DMSP platforms are stabilized using four gyroscopes (three axis stabilization), and platform orientation is adjusted using a star mapper, an Earth limb sensor, and a solar detector (Elvidge et al.\ 1997a). Daily radar bevel vector sightings of the satellites provided by Naval Space Command allow Air Force orbital mechanics models to compute geodetic subtrack of each orbit giving the position of satellites every 0.4208 seconds. This position together with OLS scan angle equations, an oblate ellipsoid Earth sea level model and digital terrain elevations from U.S. Geological Survey, EROS Data center, allow geolocation algorithms to estimate latitude and longitude for each  pixel center.

Normally the OLS is operated at high gain settings for cloud detection.  City lights are saturated in these data and radiances can not be extracted.  In 1996 and 1997, NGDC made special requests to the Air Force for collection of OLS PMT data at reduced gain settings.  This request was granted for the darkest nights of lunar cycles in March of 1996 (8 nights) and January and February (10 nights each) of 1997.  During these experimental data collections on board algorithms which adjust the visible band gain were disabled.  The major on-board algorithm which effects the night time visible band data is the  along scan gain control (ASGC), which adjust the gain dynamically in response to scene brightness.  There is also a on-board bi-directional reflectance (BRDF) algorithm designed to reduce the brightness of the image "hot-spot" which occurs where the observation angles matches the illumination angle.  The BRDF algorithm has minimal effect when lunar illumination is low, as was the case during the 28 nights in which the gain controlled data were acquired.  The two on-board gain control algorithms were turned off to simplify the retrieval of radiances from the special data collections.  

During the special data acquisition the OLS VDGA gain, which is normally operated at 60 dB, was reduced to avoid saturation in urban centres.  On one set of nights the gain was operated at a setting of 24 dB.  This produced data that avoided saturation on major on major urban centres, but did not permit detection of city edges and lighting in smaller towns.  To overcome this dynamic range limitation, data was also acquired at gain settings of 40 and 50 dB. 

\subsection{Data reduction}
\label{s2}

These data were used to assemble a cloud-free composite image calibrated to top-of-atmosphere (TOA) radiances.  The composition provides additional advantages in the removal of ephemeral lights sources, such as fire and lightning,  plus the retrieval of lights from small towns that are near the detection limits of the sensor and processing algorithms.  The primary processing steps include:
1) establishment of a reference grid with finer spatial resolution than the input imagery using the one kilometre equal area Interrupted Homolosine Projection (Goode 1925; Steinwand 1993) developed for the NASA-USGS Global 1km AVHRR project;
2) identification of the cloud free section of each orbit based on OLS-TIR data;
3) identification of lights, removal of noise and solar glare, cleaning of defective scan lines and cosmic rays;
4) projection of the lights from cloud-free areas from each orbit into the reference grid;
5) calibration to radiance units using prior to launch calibration of digital number for given input telescope illuminance and VDGA gain settings in simulated space conditions;
6) tallying of the total number of light detections in each grid cell and calculation of the average radiance value;
7) filtering images based on frequency of detection to remove ephemeral events.

The final image was transformed  into latitude/longitude projection with 30''$\times$30'' pixel size with data in  8-bit byte format and power law scaling (see Elvidge et al.\ 1999 for details). The maps of Europe described below were obtained from an image extracted from the global image. The image of Europe is 4800$\times$4800 pixels in size, starting at longitude 10\degr 30' west and   latitude 72\degr north. Orbits involved  are listed in  appendix \ref{tab1}.

Radiance range of final composite image goes from a minimum of  $1.54\times 10^{-9}$ to a maximum of   $3.17\times10^{-7}$ W cm$^{-2}$ sr$^{-1}$ $\mu$m$^{-1}$. The minimum luminance detectable for light with effective wavelength of 550 nm (which power is  $1.47\times10^{-3}$ W lm$^{-1}$)  is approximately 3 mcd m$^{-2}$. Assuming an average vertical extinction of $\sim$0.3 mag in visual band, the minimum detectable luminance on the ground  would be of the order of 4 kcd/km$^{2}$. Two un-shielded fixtures of the "globe" kind, with clean transparent glass (fixture efficiency $\sim$80 per cent), equipped with a 250 W high pressure sodium lamp with 125 lm W$^{-1}$ efficiency, and placed every square kilometer, could be sufficient to produce this luminance.

\subsection{Data analysis}
\label{s3}

\begin{figure}
\epsfysize=8.5cm % fix the y-dimension and scales x-dim. to y-dim.
\hspace{-0.5cm}\epsfbox{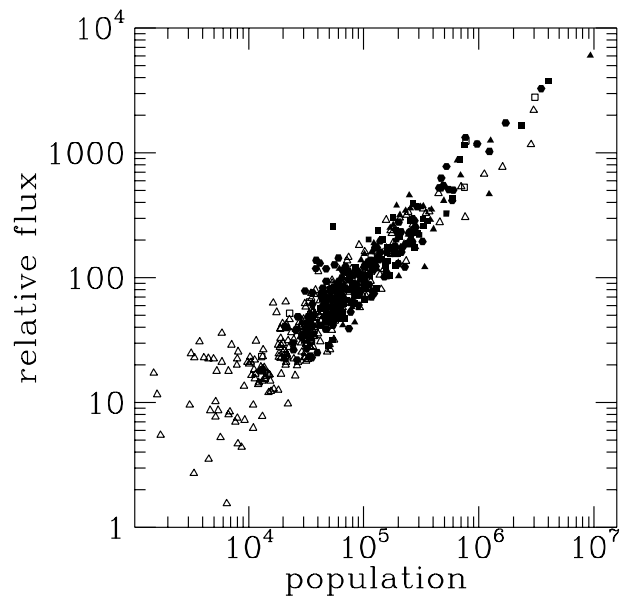} %for centering: act on hspace argument 
\caption[h]{Normalized upward flux versus city population relationship for Italy (open triangles), France (filled triangles), Germany (filled circles), Spain (filled squares), Greece (open squares). }
\label{pop2}
\end{figure}

We first analysed the composite image in order to evaluate the emission versus population relationship. We chose a number of European cities of various populations\footnote{Population data refer to 1991 for Italy, Spain and Greece, 1990 for France, 1996 for Germany and have been provided by their national bureaux of statistic.}  and measured their relative upward flux per unit solid angle summing the counts of all pixels pertinent to each city and multiplying for pixel size at that latitude. 
Fig.  \ref{pop2} show the measured upward flux of a sample of European cities normalized, for display purposes,  to the average flux of a city of $\sim10^{5}$ inhabitants in the same country.
The upward emission increases  linearly with the population.
One possible source of error is the on-board averaging of the 5$\times$5 pixel blocks during the smoothing process. During smoothing it is possible for saturated pixels in the cores of urban centres to be averaged in with unsaturated pixels to produce an unsaturated smoothed pixel value. This phenomenon will be addressed through the inclusion of OLS data acquired at even lower gain settings in the updated nighttime lights map NGDC is preparing for the 1999-2000 time period. Another uncertainty in the city analysis is that we did not precisely match the outline of the lights to match the population data reporting area.

Even if the upward emission vs. city population relationship depends on the local lighting habits, our results are in agreement with relations successfully applied or measured in other countries. Elvidge et al. \cite{e99} found a linear relation between composite radiance and population for a sample of cities of North America and obtained a smaller scatter when correlating radiance with electric consumption.
Walker \cite{w77} found  linear proportionality  between population and street light emission for a number of California cities, with a few departures above or below the mean depending on the industrial or residential character of the city. He also measured the sky illumination produced by three cities, obtaining a dependence on the power of 0.8 of their population. 
Bertiau et al. \cite{bert} used successfully a linear model with brightness proportional to population to predict light pollution, and so did Walker (1970, 1973)  and Garstang  (1987, 1988, 1989a, 1989b, 1989c, 1991a, 1991b, 1991c, 1992, 1993, 2000) who obtained a good fit with many observations including Walker \cite{w77} population - distance data.
Berry (1976) fitted well observations of sky brightness in city centers in Ontario with a propagation law for light pollution based on the approach of Treanor \cite{tre73} but assuming a dependence on the power of 0.5 of the population. However Garstang's linear models fit Berry's \cite{berry} observations well, suggesting that the power of 0.5  found by him was produced by the extinction of light emitted by outskirts of large cities in propagating to the center and does not depend on the upward flux versus population relationship (Garstang 1989a). 

Bertiau et al. \cite{bert} in the early 1970's found that the city upward emission in Italy depended on its economic and commercial development, so they were forced to include in their model a development factor. Twentyfive years later we are unable to identify an affluence effect on the brightness of italian cities. Cities in southern Italy have the same light output as comparable size cities in northern Italy, even if the former have a  per capita income that is nearly half than the latter.

The proportionality between satellite data and population would allow us to replace  the population distribution from census data, usually adopted in computations of night sky brightness, with satellite data, independently by the fact that we would be observing the light coming from the external night-time lighting or whatever else. This replacement constitutes an improvement because census data are based on city lists which, though they can be associate with the city geographical position, do not provide spatially explicit detail of the population geographical distribution.

\section{Mapping technique}
\label{s4}

Scattering from atmospheric particles and molecules spreads the light emitted upward by the sources. If $e(x,\,y)$ is the upward emission per unit area in $(x,\,y)$, the total artificial sky brightness in a given direction of the sky in a site in  $(x',\,y')$ is:
\begin{equation}
\label{int1}
b(x',\,y')=\int\int e(x,y) f((x,y),(x',\,y'))~dx ~dy\, ,
\end{equation}
where $f((x,\,y),(x',\,y'))$ give the artificial sky brightness per unit of upward light emission produced by the unitary area in $(x,\,y)$ in the site in $(x',\,y')$. The light pollution propagation function $f$ depends in general on the geometrical disposition (altitude of the site and the area, and their mutual distance), on the atmospheric distribution of molecules and aerosols and their optical characteristics in the chosen photometric band, on the shape of the emission function of the source and on the direction of the sky observed. In some works this function has been approximated with a variety of semi-empirical propagation law like Treanor Law (Treanor 1973; Falchi \& Cinzano 2000; Cinzano et al.\ 2000), Walker Law (Walker 1973; Joseph et al.\ 1991), Berry Law (Berry 1976; Pike 1976), Garstang Law (Garstang 1991b). 

We obtained the propagation function   $f((x,y),(x',\,y'))$ for each couple of points $(x,y)$ and $(x',\,y')$   with detailed models for the light propagation in the atmosphere based on the modelling technique introduced and developed by Garstang  (1986, 1987, 1988, 1989a, 1989b, 1989c, 1991a, 1991b, 1991c, 1992, 1993, 2000)  and also applied by Cinzano (2000b, 2000c, 2000d). The models assume Rayleigh scattering by molecules and Mie scattering by aerosols and take into account extinction along light paths and Earth curvature. These models allow associating the predictions with well-defined parameters related to the aerosol content, so the atmospheric conditions at which predictions refer can be well known. Here we will describe only the main lines of the models and our specific  implementation, leaving the reader to the cited papers for details.

A telescope of area $\pi d^{2}/4$ situated in the observing site O
collects from within an infinitesimal section $dV=\left( \pi \epsilon^2 u^{2} du \right)$ 
of the cone of angle $2\epsilon$ around the
line-of-sight  at a distance $u$ and with thickness $du$, a luminous flux $d\Phi$ given
by:
\BGE
\label{gar1}
d\Phi=\frac{\pi d^{2}}{4} \frac{1}{u^{2}} M_{s}(u) \xi_{1}(u) \left( \pi \epsilon^2 u^{2} du \right) ,  
\label{eq2}                                                        
\EDE
where $M_{s}(u)$ 
is  the luminous flux
scattered in unit solid angle toward the observer from particles of aerosol and molecules inside unit volume of atmosphere at the distance $u$ along the line of sight, and 
$\xi_{1}(u)$ is the extinction  of the light along its path to the telescope.

Calling $e_{s}$ the upward flux of the source and $S=M_{s}/e_{s}$ the scattered flux per unit solid angle per unit upward flux,
the propagation function $f$, expressed as total flux per unit area of the telescope
per unit solid angle per unit  total upward light emission, is found integrating eq. (\ref{eq2}) from the  site to  infinity:
\BGE
\label{gar2}
f=\int^{\infty}_{u_{0}} S(u) \xi_{1}(u) du\, ,     
\EDE
with:
\begin{equation}
\label{xi1}
 \xi_{1}= \exp \left[
-\int^{u}_{0} \left( N_{\mathrm{m}}(h)\sigma_{\mathrm{m}}+N_{\mathrm{a}}(h) 
\sigma_{\mathrm{a}} \right) dx \right],
\end{equation}
where $N_{\mathrm{m}}(h)$ and $N_{\mathrm{a}}(h)$ are respectively the vertical number densities of molecules
and aerosols, $\sigma_{\mathrm{m}}$ and $\sigma_{\mathrm{a}}$ are their scattering cross
sections. The altitude $h$ depends on the integration variable $x$, on the zenith distance and azimuth of the line of sight, on the altitudes of site and source, and on their distance.

The 
luminous flux per unit solid angle per unit upward flux coming {\em directly} from the source and scattered toward the observer from unit  volume along the line of sight is:
\BGE
\label{gar3}
S_{d}(u) =\left( N_{\mathrm{m}}(h)\sigma_{\mathrm{m}}f_{\mathrm{m}}(\omega)+N_{\mathrm{a}}(h)\sigma_{\mathrm{a}}f_{\mathrm{a}}(\omega) \right)  i(\psi,s) ,    
\label{eq4}
\EDE
where   $i(\psi,s)$  is 
the direct illuminance per unit flux produced by each source on each infinitesimal volume of atmosphere along the line-of-sight of an observer and $f_{\mathrm{m}}(\omega)$, $f_{\mathrm{a}}(\omega)$ are the normalized angular scattering functions of molecular and aerosol scattering respectively.
The  scattering angle $\omega$, the emission angle $\psi$, the distance $s$ of the section from the
source and the altitude $h$ of it, depend on the altitudes of the site and the source, their distance, the  zenith distance and the azimuth  of the line-of-sight  and  the distance $u$ along the line of sight, through some geometry.

If $I(\psi)$ is the normalized light flux per unit solid angle emitted by the considered source at the zenith distance $\psi$ and $s$ is the distance between the source and the considered infinitesimal volume of atmosphere, the illuminance per unit flux is:
\begin{equation}\label{gar4}
i(\psi,s)=I(\psi) \xi_{2}/ s^{2}
\end{equation}
in the range where there is no shielding by Earth curvature and zero elsewhere.
The extinction $\xi_{2}$ along the path is:
\BGE
\label{xi2}
\xi_{2}= \exp \left[
-\int^{s}_{0} \left( N_{\mathrm{m}}(h)\sigma_{\mathrm{m}}+N_{\mathrm{a}}(h)
\sigma_{\mathrm{a}} \right) dx \right] .
\EDE

A single scattering model is not sufficient to describe the artificial sky brightness
produced by a source. In a real atmosphere
several scatterings may occur during the travel of a photon from the source to
the telescope.  The optical thickness 
$\tau=\int k dr$, where $k$ is an attenuation coefficient, determines how important
secondary and higher scattering  is. If $\tau>>1$ (thick layer) multiple
scattering is dominant. The fraction of incident radiation which has been
scattered once is $(1-\mathrm{e}^{-\tau})$ and the fraction which it is scattered
again is of order $(1-\mathrm{e}^{-\tau})^{2}$. If $(1-\mathrm{e}^{-\tau})$ is sufficiently 
small, which happens when $\tau$ is small, secondary and higher order scattering can be
neglected. In absence of aerosol the optical thickness of the atmosphere at
wavelength of 550 nm is about 0.1 (Twomey 1977). The aerosol optical
thickness can be 0.05 in cleaner regions of 
the globe, but it can grow to higher values, even depending on seasonal
changes (Garstang 1988). Then single scattering
is the major contributor to scattered radiation but secondary scattering
is not negligible. The error in neglecting third and higher order scattering can be significant
for optical thickness higher than about 0.5.
Therefore in the computation of the  total light flux $S$ that molecules and aerosols in the infinitesimal volume scatter toward the observer  we take into account light already scattered once. 
We assumed, as Garstang (1984, 1986), that the light coming to the considered infinitesimal volume along the line of sight after a scatter has approximately a direction near that of the direct light, so that the scattering angle $\omega$ in first approximation is always the same. In this case the total illuminance $S$ can be written:
\BGE
S=S_{d}\, D_{S}\, ,
\EDE
where $D_{S}$ is a correction factor which take into account the illuminance due at light already scattered once from molecules and aerosols which can be  evaluated with the approach of Treanor \cite{tre73} as extended by Garstang (1984, 1986). Details on assumptions can be found in the quoted papers.

\section{Application}
\label{s5}

In practice, we divided the surface of Europe into land areas with the same positions and dimensions as projections on the Earth of the pixels of the satellite image. We assumed each land area be source of light pollution with an upward emission  $e_{x,y}$ proportional to the  radiance measured in the corresponding pixel multiplied by the surface area. The total artificial sky brightness at the centre of each area, given by the expression (\ref{int1}), becomes:
\begin{equation}
b_{i,j}=\sum_{h}\sum_{k}  e_{h,k} f((x_{i},y_{j}),(x_{h},y_{k}))\, .
\end{equation} 
We obtained the propagation function   $f((x_{i},y_{j}),(x_{h},y_{k}))$ for each couple of points $(x_{i},y_{j})$ and $(x_{h},y_{k})$, the positions of the observing site and the polluting area,
from eq. (\ref{gar2}) after inserting eq. (\ref{gar3}) and eq. (\ref{gar4}) multiplied for the double scattering factor $D_{S}$ computed as below.
We considered every land area as a point source located in its centre except when $i=h$, $ j=k$ in which case we used a four points approximation (Abramowitz \& Stegun 1964). The resolution of the maps, depending on results from an integration over a large zone, is greater than resolution of the original images and is generally of the order of the distance between two pixel centres. However where sky brightness is dominated by contributions of nearest land areas, effects of the resolution of the original image could became relevant.

We obtained the maps for B and V photometric astronomical bands (Johnson 1955). 

\subsection{Atmospheric model}
\label{atmod}

\subsubsection{Molecular atmosphere vertical distribution}

\begin{figure}
\epsfysize=7.8cm % fix the y-dimension and scales x-dim. to y-dim.
\hspace{0.2cm}\epsfbox{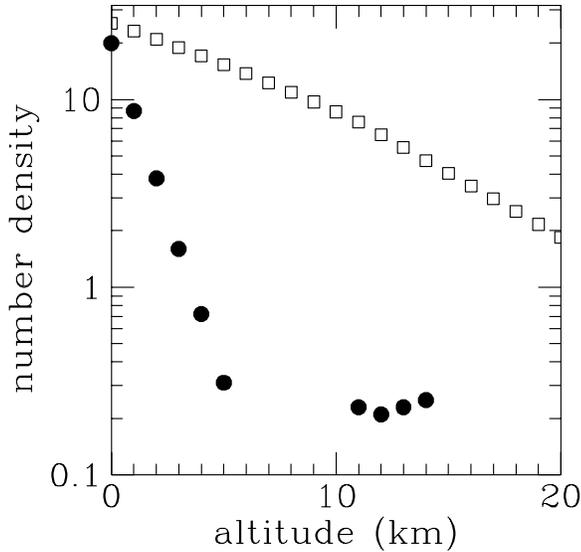} %for centering: act on hspace argument 
\caption[h]{Examples of vertical distributions of molecules (squares) and haze aerosols (circles) in the atmosphere from respectively U.S. Standard Atmosphere (1962) and Eltermann \cite{elter64} Clear Standard Atmosphere.}
\label{mol}
\end{figure}

We assumed the atmosphere in hydrostatic equilibrium under the gravitational force. Neglecting the Earth curvature, the force per unit surface supporting a molecular layer of thickness $dh$ is $dp=-g \rho dh$
where $g$ is the gravitational acceleration and $\rho$ the density of the layer.  Replacing $dp$ with the equation of state of perfect gas, working for dry air, $ dp=d\rho R\overline{T}/\overline{M}$,  where $R$ is the gas constant, $\overline{M}$ the mass of a mole of dry air and $\overline{T}$ the average temperature, and integrating, we find that in first approximation the number density $N_{\mathrm{m}}$ of the gaseous component of atmosphere decreases exponentially with altitude $h$. If $\overline{n}$ is the average number of molecules in a mole of dry air:
\begin{equation} 
N_{\mathrm{m}}(h)=\frac{\rho_{0}\overline{n}}{\overline{M}}~\exp{ \left( - \frac{\overline{M}\overline{g}}{R\overline{T}}h \right)} =
N_{\mathrm{m,0}} ~~\mathrm{e}^{-ch}\, .
\end{equation} 
Measurements show  that  this is a good approximation for the first 10 km (see e.g. Fig. \ref{mol}). We adopted this simple model as Garstang \cite{g86}, neglecting improvements done by Garstang \cite{g91a} for higher altitudes. 
As Garstang \cite{g86} we assumed an inverse scale altitude $c=\mathrm{0.104\,    km^{-1}}$ and a molecular density at sea level $N_{\mathrm{m,0}} =\mathrm{2.55\times 10^{19} cm^{-3}}$ . 

\subsubsection{Haze aerosol vertical distribution}

We are interested in average atmospheric conditions, better if typical and not for the particular conditions of a given night, so a detailed modelling of the local aerosol distribution for a given night is beyond the scope of this paper. As Garstang \cite{g86} and Joseph et al. \cite{joseph} we assumed an exponential decrease of number density for the atmospheric haze aerosols:
\begin{equation} 
N_{\mathrm{a}}(h)=N_{\mathrm{a,0}} ~~\mathrm{e}^{-ah}\, .
\end{equation} 
Measurements show (see e.g. Fig. \ref{mol}) that for the first 10 km this is a reasonable approximation. To account for presence of sporadic denser aerosol layers at various heights or at ground level as Garstang \cite{g91b} is beyond the scope of this work.  We also neglected the effects of the Ozone layer and the presence of volcanic dust studied by Garstang (1991a, 1991c). 

We take into account changes in aerosol content as Garstang \cite{g86} introducing a parameter  $K$ which measures the relative importance of aerosol and molecules for scattering light in V band:
\begin{equation}\label{kappa}
K=\frac{N_{\mathrm{a,0}} \sigma_{\mathrm{a}}}{N_{\mathrm{m,0}} \sigma_{\mathrm{m}} 11.11 \mathrm{e}^{-cH}}\,  , 
\end{equation}
where $H$ is the altitude over sea level of the ground level. 
Changing the parameter $K$ we are able to compute the map for different aerosol contents, i.e. for different products $N_{\mathrm{a}} \sigma_{\mathrm{a}}$.
As Garstang \cite{g86}, the inverse scale altitude of aerosols was assumed to be  $a=0.657+0.059 K$. Effects of changes of aerosol scale altitude has been checked in sec. \ref{calib}.
More detailed atmospheric models could be used whenever available.

\subsection{Angular scattering functions}

We take into account both Rayleigh scattering by molecules and Mie scattering by aerosols. For molecular Rayleigh scattering the angular scattering function is :
\begin{equation}\label{ray}
f_{\mathrm{m}}(\omega)= 3 (1+\cos^{2}(\omega))/16 \pi\, .
\end{equation}
The integrated  Rayleigh scattering cross section in V band was assumed to be $\sm=1.136 \times10^{-26}$ cm$^{-2}$ sr$^{-1}$ and in B band $\sm=4.6 \times10^{-27}$ cm$^{-2}$ sr$^{-1}$.

The normalized angular scattering function for atmospheric haze aerosols can be measured easily with a number of well known remote-sensing techniques like classical searchlight probing (see e.g. Elterman 1966), modern bistatic lidar probing, measurements of the day-light or moonlight sky scattering function (see e.g. Hulburt 1951; Volz 1987; Krisciunas \& Schaefer 1991).  Nevertheless in this paper we are not interested in  a specific function for a given site at a given time but in the typical average function so we adopted the function tabulated by Mc Clatchey et al. \cite{mccl} as interpolated by  Garstang \cite{g91a} and we neglected geographical gradients:
\begin{equation}
\begin{array}{l}  
\mathrm{For}\;    0\leq\omega\leq10^{\circ}: \\ 
f_{\mathrm{a}}(\omega)=7.5~ \mathrm{exp}\left(-0.1249~ \omega^{2}/(1+0.04996~ \omega^{2}) \right)\, ,\\ 
\\
\mathrm{For}\; 10^{\circ}<\omega\leq124^{\circ}: \\ 
f_{\mathrm{a}}(\omega)=1.88~ \mathrm{exp}\left( -0.07226~ \omega +0.0002406 \omega^{2} \right)\, ,\\ 
\\
\mathrm{For}\; 124^{\circ}<\omega\leq180^{\circ}:   \\ 
f_{\mathrm{a}}(\omega)=0.025 + 0.015 \sin (2.25~ \omega-369.0)\, .\\ 
\end{array} 
\end{equation}
where $\omega$ is the scattering angle in degrees.
The total integrated scattering cross section $N_{\mathrm{a}} \sigma_{\mathrm{a}}$ is given by eq. (\ref{kappa})  for a given $K$. In B band $\sigma_{\mathrm{a}}$ is  $1.216$ times the  $ \sigma_{\mathrm{a}}$ in V band.

\subsection{Upward emission function}

The normalized emission function of each area  gives the relative upward flux per unit solid angle in every direction. It is the sum of the direct emission from fixtures and the reflected emission from lighted  surfaces, normalized to its integral and is not known. 
The high number of sources contributing to the sky brightness of a site with casual distribution and orientation, smooth the shape of the average normalized emission function which can be considered in first approximation axisymmetric.
It is possible to measure the average upward emission of a chosen area at a number of different elevation angles when a large number of satellite measurements from very different orbits is available. It will be possible to obtain it directly by integrating upward emission from all lighting fixtures and all lighted surfaces on the basis of lighting engineering data and models as soon they will be available. 

In this paper we assumed that all land areas have the same average normalized emission function. This is equivalent to assuming that lighting habits are similar on average in each land area and  that differences from the average are casually distributed in the territory.
The average normalized emission function can be constrained  
from radiance measurements of cities at various distance from satellite nadir. It can also be obtained from comparison of Earth-based observations and models predictions (Cinzano, in prep.).
We chose to assume this function and check its consistency with satellite measurements, rather than directly determine it from satellite measurements because at very low elevation angles the spread is too much large to constrain adequately the function shape.

We adopted for the average normalized emission function the 
normalized city emission function from Garstang \cite{g86} :
\begin{equation}
\label{fclassic}
I(\psi)=\frac{1}{2\pi} \left[ 2a_{1} \cos \left( \psi \right) + 0.554 a_{2} \psi^{4} \right] ,
\end{equation}
where $a_{1}$ and $a_{2}$ are shape parameters. Here we assumed $a_{1}=0.46$, $a_{2}=0.54$ for the typical average function, corresponding to Garstang parameters $G=0.15$ and $F=0.15$.
This function was tested by Garstang with many comparisons between model predictions and measurements.
This author assumed this function be produced by the sum of direct emission from fixtures at high zenith distances
and Lambertian emission from lighted horizontal surfaces at higher zenith angles.
Nevertheless, upward flux can be emitted at all zenith angles both from fixtures and vertical or horizontal surfaces, so we preferred to consider Garstang's function like a parametric representation with $a_{1}$ and $a_{2}$ as shape parameters without any meaning of fraction of direct and reflected light.
We tested also the normalized city emission function of Cinzano (2000b, 2000c) which assumes a slightly higher emission at intermediate angles in respect to the function (\ref{fclassic}):
\begin{equation}
\label{fnew}
I(\psi)=\frac{1}{2\pi} \left[ a_{1} + 0.554 a_{2} \psi^{4} \right] ,
\end{equation}
assuming $a_{1}=0.79$ and $a_{2}=0.21$.
Comparison between these functions are presented by Cinzano \cite{c99a}.

We checked these functions by studying the relation between the upward flux per unit solid angle per inhabitant of a large number of cities and their distance from the satellite nadir in a single orbit satellite image taken the 13th January 1997 h20:27  from satellite F12 which was chosen for its large cloudfree area.
Taking into account the average orbital distance $R_{S}$ and the Earth curvature radius $R_{T}$, it is possible, with some geometry, to relate the  distance $D$ from satellite nadir with the emission angle $\psi$:
\begin{equation}
\label{nad1}
\psi= \frac{D}{R_{T}}\arcsin  
\frac{R_{T} \sin (D/R_{T})}
{s}\, ,
\end{equation}
with:
\begin{equation}
s=\sqrt{R_{T}^{2}+(R_{T}+R_{S})^{2}-2R_{T}(R_{T}+R_{S})\cos (\frac{D}{R_{T}})}\, .
\end{equation} 
The flux per unit solid angle per inhabitant in relative units is obtained dividing the measurements for the correspondant extinction coefficient $\xi_{3}(\psi)$ computed for curved-Earth from eq. (20) of Garstang \cite{g89a}:
\begin{equation}\label{nad2}
\begin{array}{l} 
\xi_{3}= \mathrm{exp} \left[- N_{\mathrm{m}}  \sigma_{\mathrm{m}} \left( A1+ 11.778 K A2 \right) \right] \\
A1=\frac{1}{c} \sec \psi  
\left(1- \mathrm{e}^{-cs \cos \psi} + \frac{16}{9\pi}\frac{ \tan^{2} \psi}{2cR_{T}} \, B1 \right)\\
B1= \left( c^{2}s^{2}\cos^{2}\psi + 2 cs \cos \psi +2 \right)  \mathrm{e}^{-cs \cos \psi} -2 \\
A2=\frac{1}{a} \sec \psi  
\left(1- \mathrm{e}^{-as \cos \psi} + \frac{16}{9\pi}\frac{ \tan^{2} \psi}{2aR_{T}} \, B2 \right)\\
B2= \left( a^{2}s^{2}\cos^{2}\psi + 2 as \cos \psi +2 \right)  \mathrm{e}^{-as \cos \psi} -2 \, .\\
\end{array}
\end{equation}

\begin{figure}
\epsfysize=7.8cm % fix the y-dimension and scales x-dim. to y-dim.
\hspace{0.2cm}\epsfbox{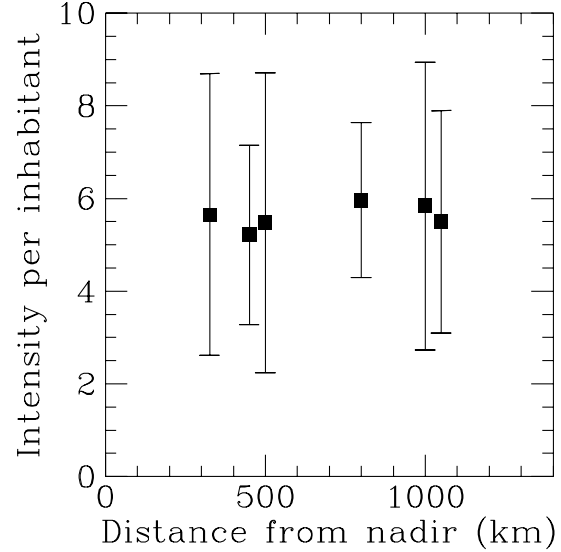} %for centering: act on hspace argument 
\caption[h]{Relative flux per unit solid angle per inhabitant averaged over ranges of 100 km measured for a sample of cities versus their distance from satellite nadir.}
\label{nadir1}
\end{figure}

Fig. \ref{nadir1} shows the relative flux per unit solid angle per inhabitant averaged over ranges of 100 km. 
\begin{figure}
\epsfysize=4.5cm % fix the y-dimension and scales x-dim. to y-dim.
\hspace{-0.1cm}\epsfbox{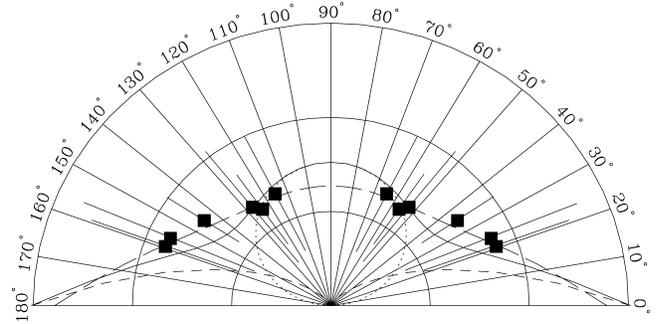} %for centering: act on hspace argument 
\caption[h]{Average upward flux per unit solid angle per inhabitant compared with Garstang upward emission function (solid line) and Cinzano upward emission function (dashed line) for the adopted shape parameters. The components of the Garstang function are also shown (dotted and short-dashed lines).}
\label{nadir2}
\end{figure}

In Fig. \ref{nadir2} we show the $I(\psi)$ obtained with eq. (\ref{nad1}) and (\ref{nad2}) compared with the Garstang \cite{g86} function (solid line) and with the Cinzano \cite{c99a} function (dashed line) with the assumed shape parameters. The fits are good with both. Errorbars are not necessarily related to fluctuations in function shape but rather with fluctuations in the total flux per inhabitant.
Given that the extinction effects seems to balance the changes in measured flux at angles off from nadir (see Fig.\ref{nadir1}), we did not need to correct the input images from single orbits for these off-nadir effects when computing the composite image.

Snow reflects approximately  60 per cent of downlight changing the shape of the upward emission function. Lighted roads usually are cleaned in a few days so reflection of street lighting by snow on road surfaces is unlikely to be important, but reflection of the artificial sky light by snow on the rest of the land could enhance noticeably the upward flux in more polluted areas. Assuming roughly that 10 per cent of the upward flux be scattered downward by atmospheric particles and molecules and that 60 per cent of it be reflected upward again by the snow covered terrain,  the increase of artificial sky brightness by snow reflexion of sky light could reach 6 per cent. The upward flux due to snow reflection in some zones of Europe (our images are taken in winter) is likely detected by the OLS-PMT, but the different shape of the upward emission function could produce small errors. We plan to use specific satellite surveys to detect snowed areas where the upward emission function must be corrected.
Even the offshore lights in the North Sea, where oil and gas production sites are active, could have a different upward emission function.

\subsection{Geometric relations}

In this paper we are more interested in understanding and comparing light pollution distributions in the European territory rather than in predicting the effective sky brightness for observational purposes, so we  computed the artificial sky brightness at sea level, in order to avoid introduction of altitude effects in our maps.
We plan to take account of altitudes  in a forecoming paper devoted to mapping the naked-eye star visibility  which requires the computation of star-light extinction and natural sky brightness for the altitude of each land area.

With the hypothesis of  sea level, geometrical relations from Garstang \cite{g89a}   between quantities shown in Fig. \ref{geo} taking into account Earth curvature, became simpler. They are listed in the appendix \ref{apb}. 
  In eq. (\ref{kappa}) now is $H = 0$.
\begin{figure}
\epsfysize=8cm % fix the y-dimension and scales x-dim. to y-dim.
\hspace{.5cm}\epsfbox{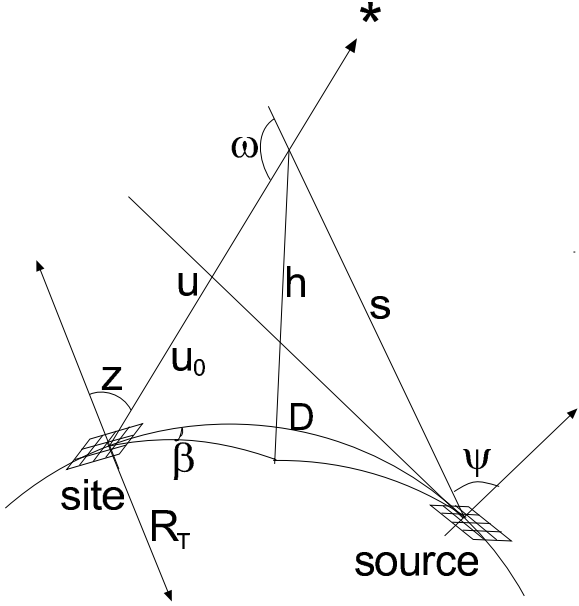} %for centering: act on hspace argument 
\caption[h]{Geometrical relationships.}
\label{geo}
\end{figure}
Equations (\ref{xi1}), (\ref{xi2}) have been integrated by Garstang (1989a: eq. (18), (19), (22) and eq. (20), (21)). They are listed in appendix \ref{apb} too. 

Correction for double scattering of Garstang \cite{g89a} became:
\begin{equation}\label{ds}
D_{S}= 1\! +  N_{\mathrm{m,0}} \sigma_{\mathrm{m}} (11.11 Kf_{2}+\frac{f_{1}}{3})\, ,
\end{equation}
where $f_{1}$ and $f_{2}$ are given in eq. (\ref{effe}).
Integration of eq. (\ref{gar2}) must be done only where the Earth curvature does not shield the line of sight from the source. We need to  start integration from $u_{0}$ as given by  (from  eq. (12) and (13) of Garstang 1989a):
\begin{equation}
u_{0}=\frac{2R_{T} \sin^{2}(D/2R_{T})}{\sin z \cos \beta \sin(D/R_{T})+ \cos z \cos(D/R_{T})}\, .
\end{equation}

We neglected the presence of mountains which might shield the light emitted from the sources from a fraction of the atmospheric particles along the line-of-sight of the observer.  
Assuming flat Earth, mountains  between source and observatory  shield the light emitted from the source with an angle less than  $\theta =
arctg\frac{H}{p}$ where  $H$ is the height of the mountain and $p$ its
distance from the source. The ratio between the
luminance in the shielded and not shielded cases is given, in first approximation, by the ratio between the number of
particles  illuminated in the two cases: 
\begin{equation}  \frac{b_s}{b_{ns}}\approx\frac{  
 \sigma_{\mathrm{a}} f_{\mathrm{a}}(\omega) \int^{\infty}_{\frac{Hq}{p}} N_{\mathrm{a}}(h) dh  +  
\sigma_{\mathrm{m}} f_{\mathrm{m}}(\omega) \int^{\infty}_{\frac{Hq}{p}} N_{\mathrm{m}}(h) dh    
}  
{  
 \sigma_{\mathrm{a}} f_{a}(\omega) \int^{\infty}_{0} N_{\mathrm{a}}(h) dh +  
 \sigma_{\mathrm{m}} f_{m}(\omega) \int^{\infty}_{0} N_{\mathrm{m}}(h) dh
}  \, ,
\end{equation}
where $q$ is the distance of the site from the source, 
$N_{\mathrm{m}}(h)$ and $N_{\mathrm{a}}(h)$ are respectively the vertical number densities of molecules
and aerosols at the altitude $h$, $\sigma_{\mathrm{m}}$ and $\sigma_{\mathrm{a}}$ are their scattering
sections  and roughly $\omega \approx \pi-\theta$.
This expression shows that,  given the vertical extent of the atmosphere in respect to the highness of the mountains, the shielding is not negligible only when the source is very near the mountain and both are quite far from the site (Garstang 1989a; see also Cinzano 2000c): $ \frac{q}{p} >> \frac{h_{\mathrm{a}}}{H}$ where $h_{\mathrm{a}}$ is the vertical scale height of the atmospheric aerosols.  This condition in the considered territories usually applies to poorly lighted areas only, which produce little pollution. Earth curvature emphasizes this behaviour.

\subsection{Relation with atmospheric conditions}

The adopted modelling technique allows us to assess the atmospheric conditions for which a map is computed giving observable quantities like the vertical extinction at sea level  in magnitudes (Garstang 1989a):
\begin{equation}\label{vertext}
\Delta m= 1.0857 N_{\mathrm{m,0}} \sigma_{\mathrm{m}}\!  \left( \left( \frac{550}{\lambda}\right)^{4} \frac{1}{c}+ 
\left( \frac{550}{\lambda}\right) \frac{11.778 K}{a} \right),
\end{equation}
where for V band $\lambda=550$ and for B band $\lambda=440$.
Neglecting Earth curvature,
the horizontal visibility defined as the distance at which a black object would show a brightness of 0.98 of the brightness of background horizon due to scattered light is (Garstang 1986):
\begin{equation}\label{orvis}
\Delta x=  \frac{3.91}{N_{\mathrm{m,0}} \sigma_{\mathrm{m}}}
\left( \left( \frac{550}{\lambda}\right)^{4}+
11.778 K \left( \frac{550}{\lambda}\right) \right)^{-1} .
\end{equation}
Effects of curvature have been discussed by the cited author.
Other relations exist with measurables like the Linke turbidity factor for total solar radiation received on Earth (Garstang 1988).
The optical thickness $\tau$ is (from eq. (22) of Garstang 1986):
\BGE
\label{tau}
\tau=\Delta m / 1.0857\, .
\EDE
With $K$=1 we obtain $\tau=0.3$ so double scattering approximation is adequate in our map computations. In order that $\tau \le 0.5$ we need $K\le$2.2.

\subsection{Spectral emission}

DMSP satellites do not carry a spectrograph able to obtain spectra of the upward light emitted by each land area. However it is possible to recover information about the spectral emission from the differential effects of extinction on light of different wavelengths.
Calling $I(\lambda)$ the specific radiance of each land area at wavelength $\lambda$, $T(\lambda)$ the PMT sensitivity curve and  $\xi_{4}(\lambda,K)$ the vertical extinction for a given clarity parameter $K$, the radiance $r(\lambda)$ measured by the OLS-PMT  is:
\begin{equation}
\label{laplace}
r(K)=\int_{0}^{\infty} T(\lambda)I(\lambda)\xi_{4}(\lambda,K) \,d\lambda\, .
\end{equation}
The vertical extinction can be obtained from eq. (\ref{vertext}):
\begin{equation}
\begin{array}{l}
\xi_{4}(\lambda,K)= \exp \left( p_{1}\lambda^{-4}\right) \cdot \exp \left( p_{2}K\lambda^{-1} \right)\\
p_{1}= - N_{\mathrm{m,0}}  \sigma_{\mathrm{m}} \lambda_{0}^{4}/c\\
p_{2}= - 11.778\,N_{\mathrm{m,0}}  \sigma_{\mathrm{m}} \lambda_{0}/a\, ,\\
\end{array}
\end{equation}
where the molecular scattering cross section $\sigma_{\mathrm{m}}$ at $\lambda_{0}=550$ nm, the molecular density at sea level $N_{\mathrm{m,0}}$ and the inverse scale heights of molecules and aerosols $c$  and  $a$ are given in section \ref{atmod}. 
Equation (\ref{laplace}) could be inverted with a Richardson-Lucy algorithm  in order to recover $I(\lambda)$ from $r(K)$, a function which can be obtained comparing many radiance calibrated images obtained in different atmospheric conditions. The average clarity parameter $K$ for each image can be obtained comparing the variations of measured radiance with increasing distance from satellite nadir.

\begin{figure}
\epsfysize=16cm % fix the y-dimension and scales x-dim. to y-dim.
\hspace{0cm}\epsfbox{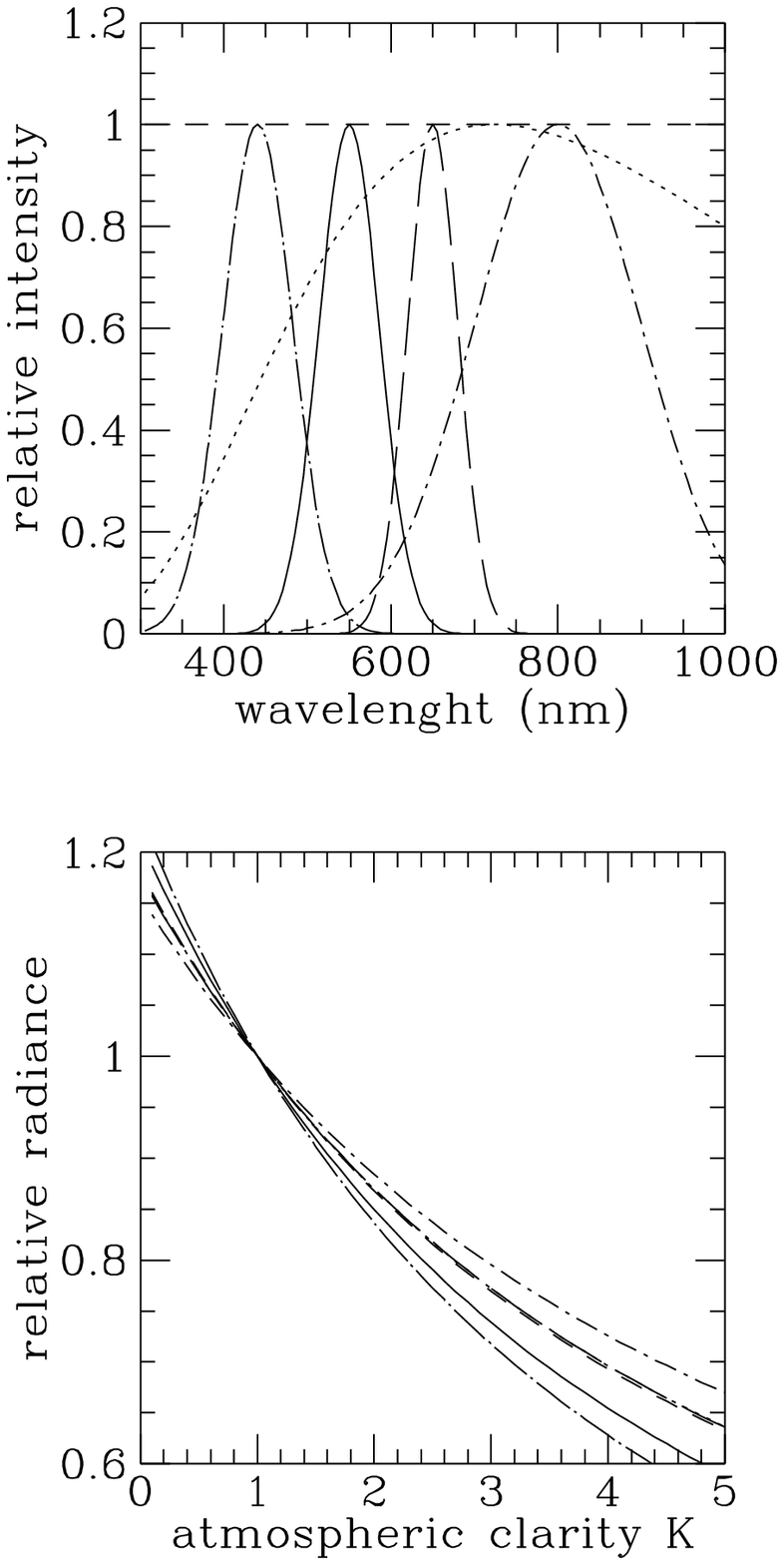} %for centering: act on hspace argument 
\caption[h]{The $r(K)$ curves (lower panel) for some example spectra (upper panel). Spectra with a similar effective wavelength after convolution for the PMT sensitivity curve, like e.g. a constant (dashed curve), a blackbody at 4000\degr K (dotted curve) and a narrow gaussian centered at 650 nm (long dashed curve),  give almost the same $r(K)$ curve. However spectra with lower or higher effective wavelength, like a gaussian centered at 550 nm (solid curve) or at 800 nm (dot-dashed curve), give $r(K)$ curves with respectively more or less pendency.}
\label{spectra}
\end{figure}

Fig. \ref{spectra} shows in the lower panel the $r(K)$ curves produced by the example spectra $I(\lambda)$ in the upper panel. Curves have been scaled so that $r(K\!\!  =\!\!  1)=1$. Spectra with a similar effective wavelength after convolution with the PMT sensitivity curve, like e.g. a constant (dashed curve), a blackbody at 4000\degr K (dotted curve) and a narrow gaussian centered at 650 nm (long dashed curve),  give almost the same $r(K)$ curve. However spectra with lower or higher effective wavelength, like a gaussian centered at 440 nm (dot-long dashed curve), 550 nm (solid curve) or at 800 nm (dot-dashed curve), give different $r(K)$ curves.

We obtained the maps for B and V photometric astronomical bands. The relative B-V color index of each land area could be inferred from the $r(K)$ curves but requires some assumptions on the shape of the spectra. In order to be simple in this paper we assumed it to be constant everywhere and we only took into account the different propagation of the light in the atmosphere in the two bands. We plan to study differences in city spectra in future works.

\subsection{Calibration}

We calibrated the maps on the basis of both (i) accurate measurements of sky brightness together with extinction from the Earth-surface and (ii) analysis of before-fly radiance calibration of  OLS-PMT.

\subsubsection{ Calibration with Earth-based  measurements} 
\label{calib}

A detailed calibration requires sky brightness measurements  at a large number of  sites on  sea level, taken in nights with the same vertical extinction and horizontal visibility of the map under calibration, averaged over many nights in order to smooth atmospheric fluctuations. Observations need to be {\em under  the atmosphere}, i.e. as actually observed from the ground without any extinction correction applied.
To obtain the artificial sky brightness it would be necessary to measure the natural sky brightness in some sites where the maps suggest that the artificial one is negligible, and subtract the mean from all measurements. Moreover given the fast growth rate of artificial sky brightness, which e.g. in Italy reaches 10 per cent per year (Cinzano 2000d), measurements have to be taken in the same period and the calibration will refer to that time.

Measurements of sky brightness in Europe in V and B bands  in the period 1996-1999 are scarce, so we calibrated our maps with all available measurements in the chosen bands taken in 1998 and in 1999 in clean or photometric nights even if extinction was not available or not exactly the required one (Catanzaro \& Catalano 2000; Cinzano 2000d; Della Prugna 1999; Favero et al.\ 2000; Piersimoni, Di Paolantonio \& Brocato 2000; Poretti \& Scardia 2000; Zitelli 2000). Some data has been taken by one of us for this purpose with a small portable telescope and a CCD device (Falchi 1999). Most of the sites are at sea level but we included also a few sites at altitudes under 1300 m o. s. l.. In lack of measurements of natural sky brightness in Europe at sea level, we assumed it at minimum solar activity  $B=22.7$ mag arcsec$^{-1}$ in B band, and $V=21.6$ mag arcsec$^{-1}$ in V band, estimating an incertitude of at least $\pm0.1$ mag arcsec$^{-1}$. Natural sky brightness increases when solar activity increases (Walker 1988) and the solar activity in 1998 was close to minimum but not at the minimum, so it could be underestimated and consequently the artificial brightness in darker sites considerably overestimated. The sky brightness has been transformed into photon radiance with formulae of Garstang (1989a: eq. (28) and eq. (39)).

A least square fit of a straight line $y=a+x$ over the logarithmic measured radiances versus the logarithmic predicted radiances gives the logarithmic calibration coefficients $a_{B}=-0.63\pm0.04$ and $a_{V}=0.00\pm0.04$.  We assumed unavailable the incertitudes of measurements as given by atmospheric conditions and emission function fluctuations. The uncertainty of the calibration coefficients produces an incertitude  of  about 10 per cent in the calibrated predicted radiances. However single data points show differences even of about 60 per cent, so that we consider safer to assume this last value as estimate of the uncertainty of our calibrated predictions for a given site. More precise calibrations will be possible when a  large number of measurements of sky brightness at sea level together with extinction become available. A large CCD measurement campaign is being set up.

\begin{figure}
\epsfysize=9.2cm % fix the y-dimension and scales x-dim. to y-dim.
\hspace{-0.5cm}\epsfbox{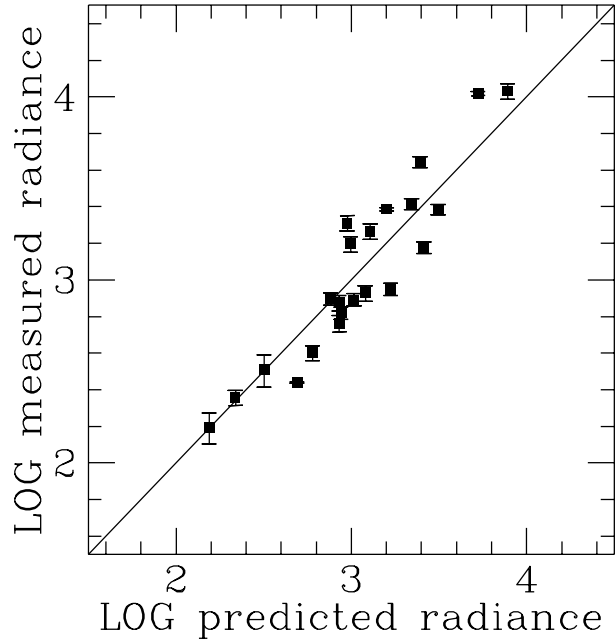} %for centering: act on hspace argument 
\caption[h]{Measurements of artificial sky brightness versus calibrated map predictions  in V band.}
\label{calv}
\end{figure}

\begin{figure}
\epsfysize=8.9cm % fix the y-dimension and scales x-dim. to y-dim.
\hspace{-0.5cm}\epsfbox{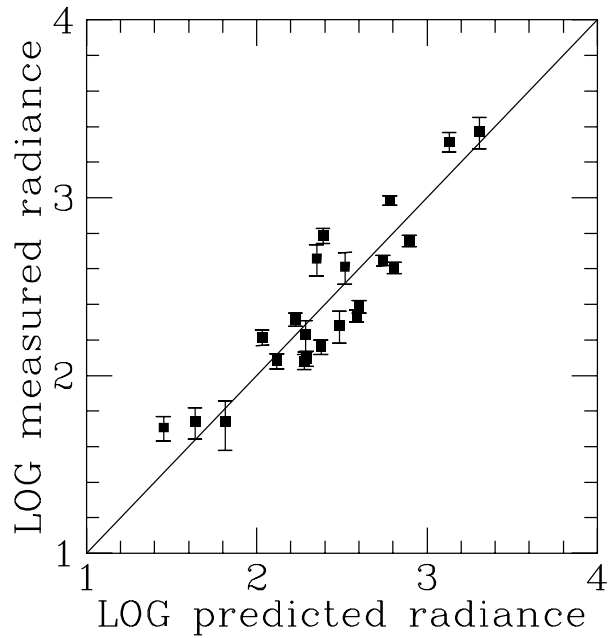} %for centering: act on hspace argument 
\caption[h]{Measurements of artificial sky brightness versus calibrated map predictions in B band.}
\label{calb}
\end{figure}

In Figs \ref{calv} and \ref{calb} we compared our calibrated predictions with the available measurements of artificial sky brightness respectively in V and in B bands. Photon radiance in V band is expressed in units of 0.3419 10$^{10}$ ph s$^{-1}$ m$^{-2}$ sr$^{-1}$, corresponding approximately to a luminance of one $\mu$cd m$^{-2}$, and in B band in units of 10$^{10}$ ph s$^{-1}$ m$^{-2}$ sr$^{-1}$. Errorbars indicate measurement errors which are much smaller than the effects of fluctuations in atmospheric conditions.

We checked the effects on predictions of Figs \ref{calv} and \ref{calb} of changes of (i) atmospheric conditions and (ii) the shape of $I(\psi)$.
An increase of the aerosol content parameter $K$ increases exponentially the extinction and linearly the scattering along the line of sight, 
so that, excluding a small area very near the source where the extinction is negligible, the predicted artificial sky brightness decreases everywhere: a well known result (Garstang 1986, 1989a, 2000; Cinzano 2000c).
The decrease is larger for areas farther  from sources. These usually are darker, so in the log-log diagram, increasing haze  darker sites move toward low predicted brightness more than less dark sites. A change of aerosol scale length has a similar effect.
A change of the shape of the upward emission function $I(\psi)$ has a different effect. Decreasing the relative emission at  low elevations (large $\psi$) the sky brightness is decreased approximately proportionally in almost all sites, excepts very near the source (see also Cinzano \& Diaz Castro 2000), so in a log-log diagram all sites move toward low predicted brightness of nearly the same quantity.
This opens a way to detect cities with more light-wasting installations. In fact if a series of measurements of sky brightness at increasing distances from a city lies on a straight line parallel to the calibration line but displaced  toward lower predicted values, likely the relative emission of the city at low elevation is higher than assumed in making the map. This is a typical result of poorly shielded or too much inclined street light.

Shifts in measurements obtained with different instrumental setups could arise because, as will be shown in Fig. \ref{prefly}, the average emission spectrum has typically its maximum at a side of the V band sensitivity curve, so that the resulting measurements are quite sensitive to little differences between  the instrumental curve and the standard V band curve.

\subsubsection{Calibration with pre-fly irradiance measurements}
\label{preflycal}

Map calibration based on pre-fly irradiance calibration of OLS PMT requires the knowledge, for each land area $(i,\,j)$, of (i) the average vertical extinction $\Delta m$ during satellite observations and (ii) the relation between the radiance in the chosen photometrical band and the radiance measured in the PMT spectral sensitivity range, which depends on the emission spectra. Both of them are unknown. 

If $\overline{r}$ is the energetic radiance in $\left[\mathrm{10^{10} W cm^{-2} sr^{-1} }\right]$,  the upward energy flux in $\left[W\right]$ in the PMT photometric band  is:
\begin{equation}
\label{prefcab}
e(i,j)  = \overline{r}(i,j)  \frac{A(i,j) }{I(\psi\! =\! 0)} \, 10^{0.4 \Delta m}\, ,
\end{equation}
where $I(\psi)$ is the upward emission function and A is the surface area in $km^{2}$ of the land area $(i,\,j)$:
\begin{equation}
A(i,j)=\left( \frac{2 \pi \Delta x}{360 \cdot 60 \cdot  60}  R_{T}\right)^{2} \cos(l)\, ,
\end{equation}
with $l$  latitude of the land area, $R_{T}$  average Earth radius in km and $\Delta x$ pixel size in arcsec. 

The photon radiance in the V photometric band corresponding to the energetic radiance measured by the PMT is:
\begin{equation}
\label{photons}
\begin{array}{l}
\overline{f_{r}}(i,j) =\frac{\int^{\infty}_{0} T_{\mathrm{V}}(\lambda) \,I(\lambda)\frac{\lambda}{h c} \, d\lambda}
{\int^{\infty}_{0} T_{\mathrm{P\! M\! T}}(\lambda)\, I(\lambda) \,\frac{\lambda}{h c}\, d\lambda}\,
\frac{\int^{\infty}_{0} T_{\mathrm{P\! M\! T}}(\lambda) \,S(\lambda)\frac{\lambda}{h c} \, d\lambda}
{\int^{\infty}_{0} T_{\mathrm{P\! M\! T}}(\lambda)\, S(\lambda) \, d\lambda}\,\overline{r}(i,j) \\
=\frac{\int^{\infty}_{0} T_{\mathrm{V}}(\lambda) \,I(\lambda)\frac{\lambda}{h c} \, d\lambda}
{\int^{\infty}_{0} T_{\mathrm{P\! M\! T}}(\lambda)\, I(\lambda) \,\frac{\lambda}{h c}\, d\lambda}\,
\frac{<\lambda>}
{hc}\, \overline{r}(i,j) \, ,
\end{array}
\end{equation}
where $T_{V}$ and $T_{\mathrm{P\! M\! T}}$ are the sensitivity curves respectively of V band and PMT detector, $I(\lambda)$ is the energy spectrum of the emission from the chosen land area, $S(\lambda)$ is the energy spectrum of the pre-fly calibration source, $h$ is the Plank constant and $c$ is the velocity of light. The second ratio is the effective wavelength $<\lambda>$ of the combination of the sensitivity curves of the PMT and the calibration source,  divided by $hc$. 

\begin{figure}
\epsfysize=8cm % fix the y-dimension and scales x-dim. to y-dim.
\hspace{0cm}\epsfbox{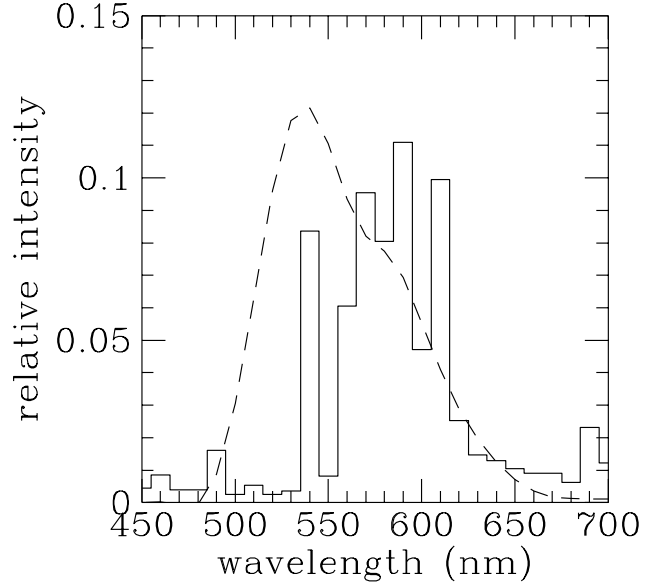} %for centering: act on hspace argument 
\caption[h]{The composite spectrum of upward emission (solid line) used for checking the consistency of Earth-based map calibration with PMT radiance calibration.  We roughly assumed 50 per cent of the total power produced by HPS  lamps  and 50 per cent by Hg vapour lamps.  Dashed line shows the V band sensitivity curve.}
\label{prefly}
\end{figure}

In order to check the consistency of our Earth-based V-band calibration  with pre-fly radiance calibration of PMT images,  we obtained a tentative map calibration
assuming for all land areas an average vertical extinction $\Delta m=0.3$ mag V at sea level and constructing a synthetic spectrum for a typical night-time lighting. We very roughly assumed  that 50 per cent of the total power emitted by each land area be produced by HPS  lamps (SON standard) and 50 per cent by Hg vapour lamps (HQI). The composite spectrum is visible in Fig. \ref{prefly} (solid line) together with the V band (dashed line). 
The result of integration of eq. (\ref{photons}) is  $\frac{\overline{f_{r}}}{\overline{r}}\approx1.48\times10^{18}$ ph s$^{-1}$  W$^{-1}$, much less than the photon flux per unit power at 550 nm, $\sim2.79\times10^{18}$ ph s$^{-1}$  W$^{-1}$.  Eq. (\ref{prefcab}) and eq. (\ref{photons}), taking in account the internal constants of the program and omitting the $\cos{(l)}$ already included in it, give the V-band logarithmic calibration coefficient $a_{V}=-0.01$. In spite of the uncertainties both in the extinction and in the average emission spectrum, this calibration agree very well with the Earth-based calibration. 

\section{Results}
\label{s6}

\begin{figure}
\epsfysize=18cm % fix the y-dimension and scales x-dim. to y-dim.
\hspace{0.4cm}\epsfbox{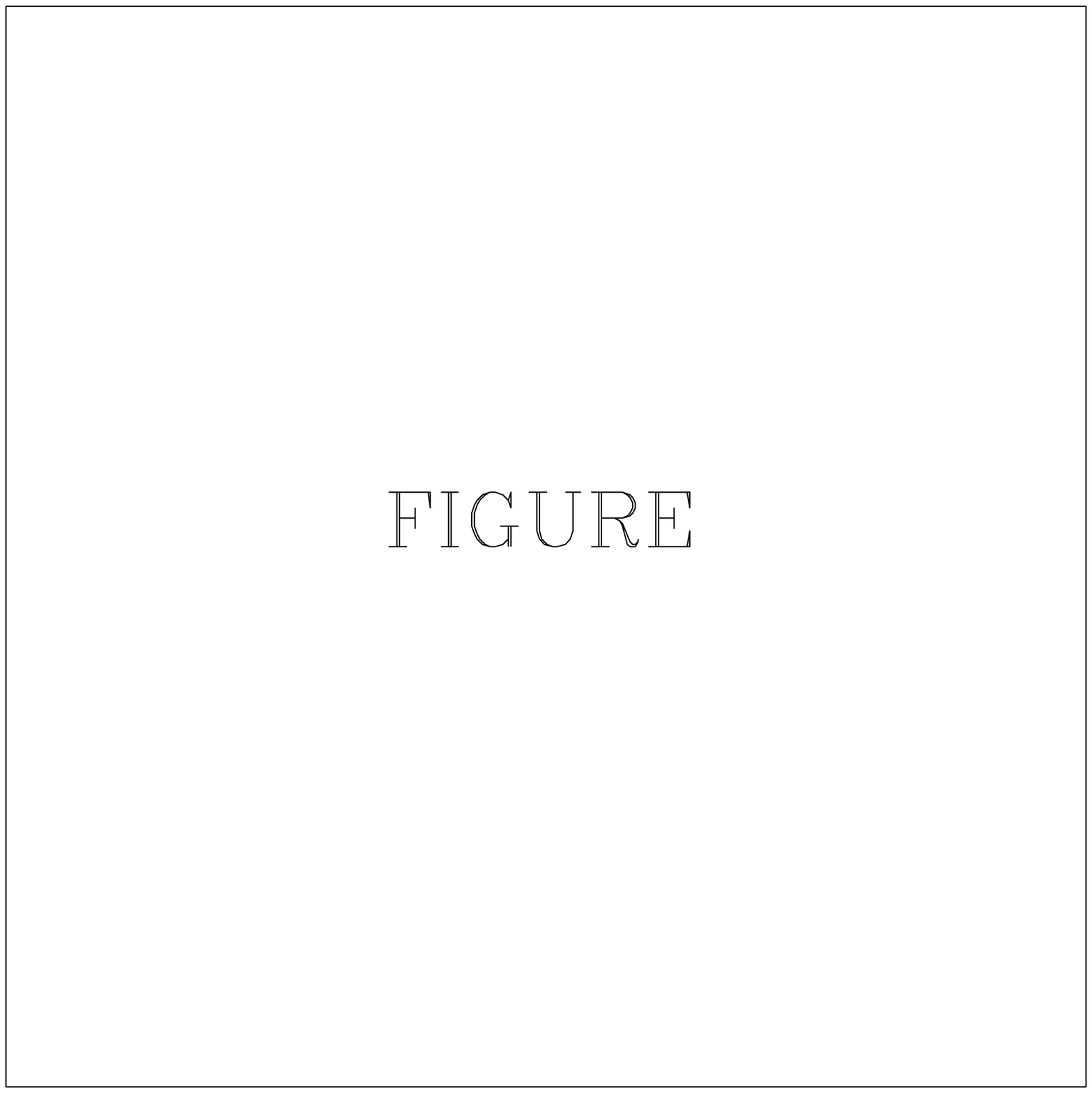} %for centering: act on hspace argument 
\caption[h]{Artificial sky brightness at sea level in Europe in V band for aerosol content parameter $K=1$.}
\label{res1}
\end{figure}

\begin{figure}
\epsfysize18cm % fix the y-dimension and scales x-dim. to y-dim.
\hspace{0.4cm}\epsfbox{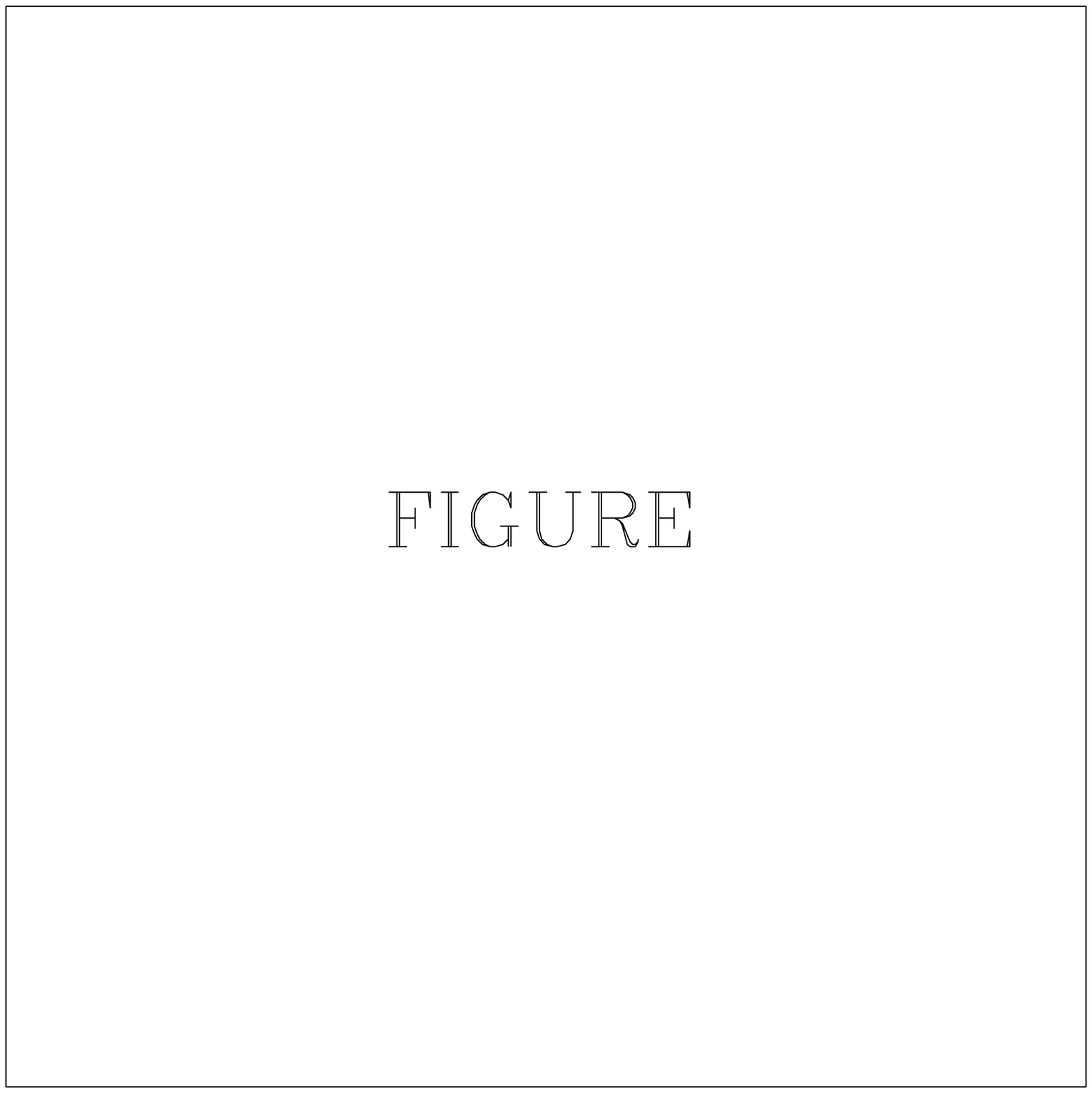} %for centering: act on hspace argument 
\caption[h]{Artificial sky brightness at sea level in Europe in V band for aerosol content parameter $K=1$.}
\label{res2}
\end{figure}

\begin{figure}
\epsfysize=18cm % fix the y-dimension and scales x-dim. to y-dim.
\hspace{0.4cm}\epsfbox{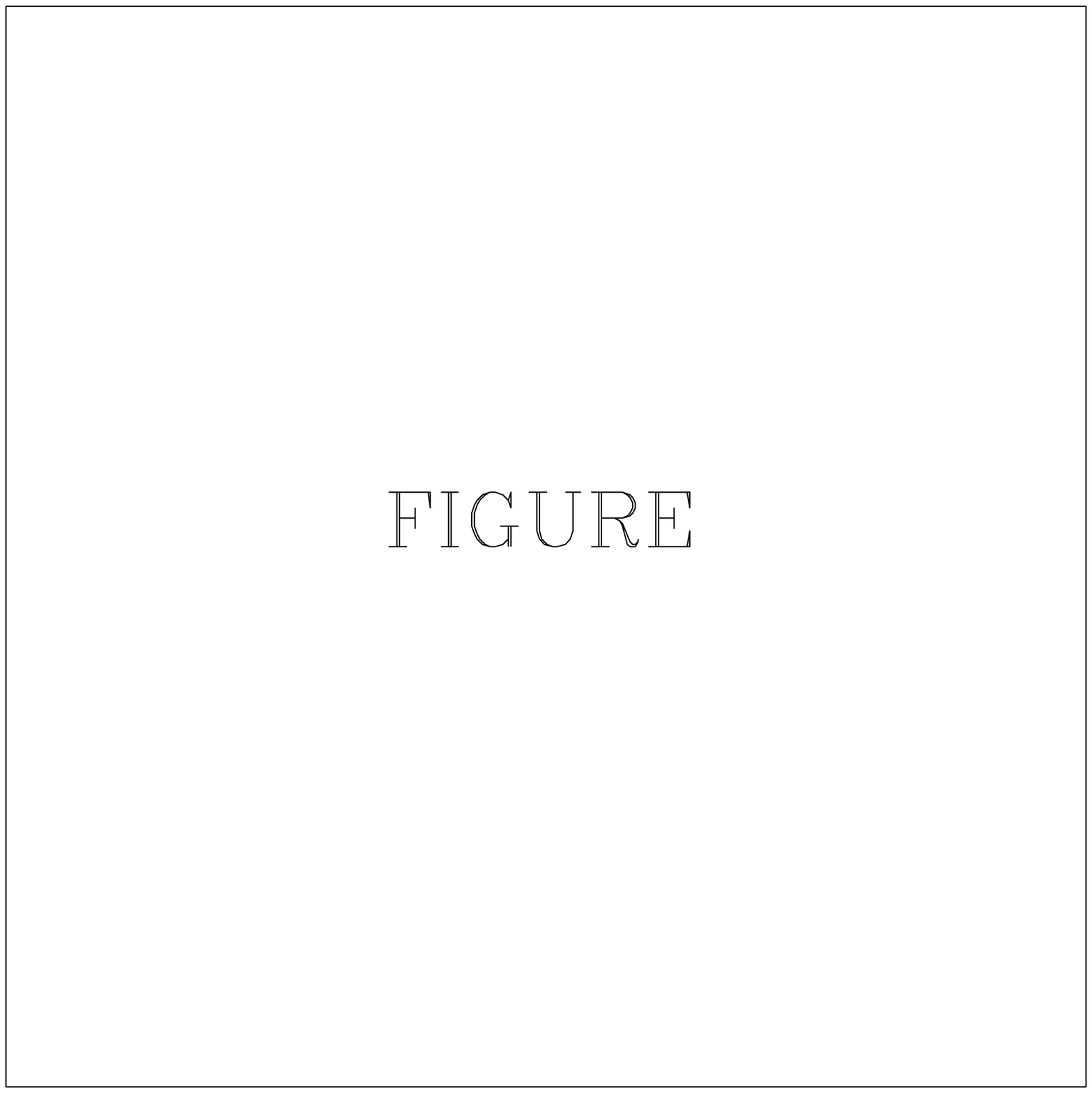} %for centering: act on hspace argument 
\caption[h]{Artificial sky brightness at sea level in Europe in B band for aerosol content parameter $K=1$.}
\label{res3}
\end{figure}

\begin{figure}
\epsfysize=18cm % fix the y-dimension and scales x-dim. to y-dim.
\hspace{0.4cm}\epsfbox{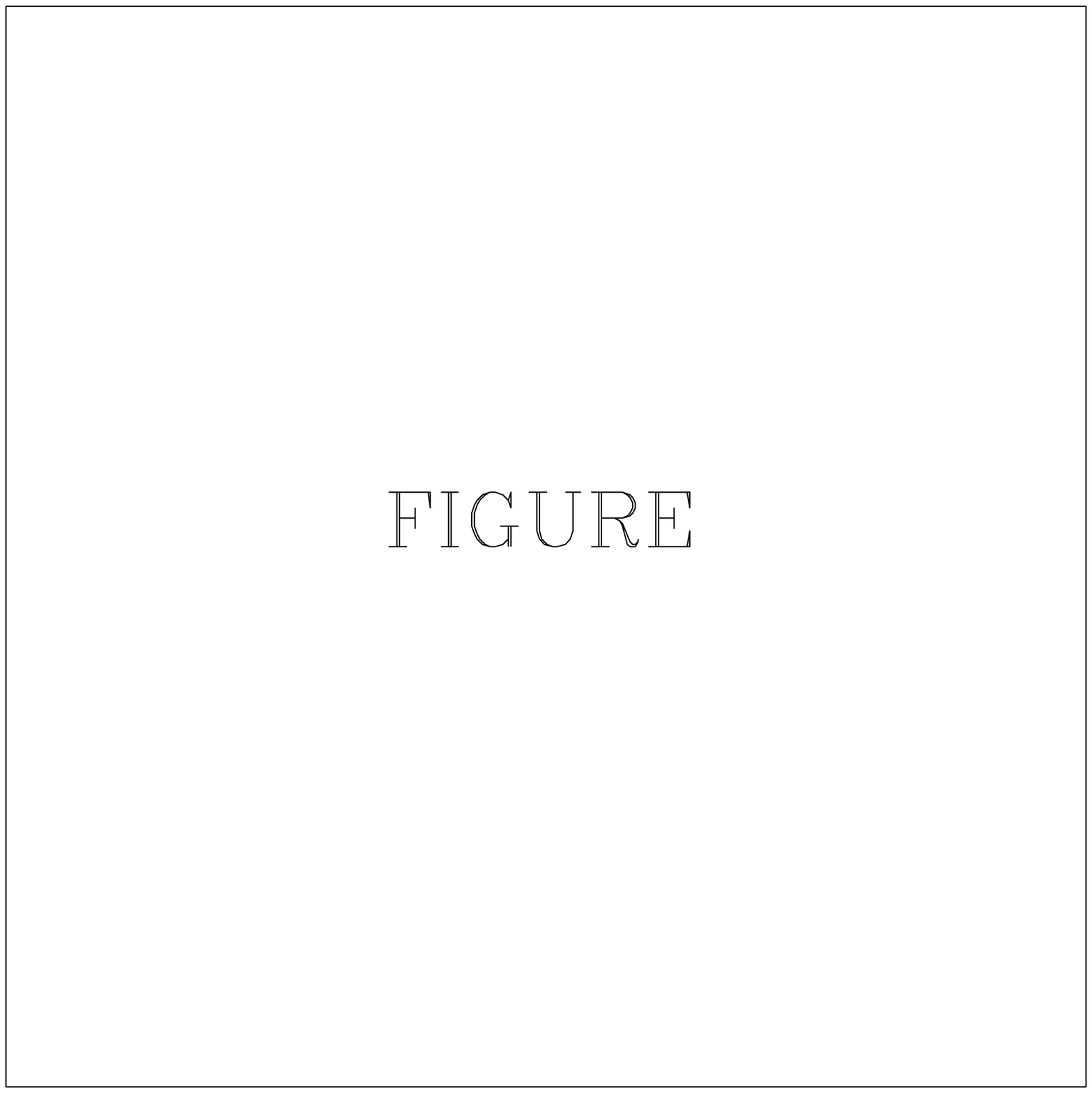} %for centering: act on hspace argument 
\caption[h]{Artificial sky brightness at sea level in Europe in B band for aerosol content parameter $K=1$.}
\label{res4}
\end{figure}

Figs \ref{res1}, \ref{res2}, \ref{res3} and \ref{res4} show the artificial sky brightness in Europe at sea level in V and B bands.  
Colours correspond to ratios between the artificial sky brightness and the natural sky brightness of: $<$0.11 (black), 0.11-0.33 (blue), 0.33-1 (green), 1-3 (yellow), 3-9 (orange), $>$9 (red). 
Original maps are 4800$\times$4800 pixel images saved  in 16-bit standard {\sc fits} format  with {\sc fitsio} Fortran-77 routines developed by HEASARC at the NASA/GSFC. Images have been analysed with {\sc ftools 4.2} analysis package by HEASARC and with {\sc quantum image 3.6} by Quantum Image Systems.  Maps have been computed for clean atmosphere with aerosol clarity $K=1$, corresponding to a vertical extinction of $\Delta m =0.33$ mag in V band, $\Delta m = 0.56$ mag in B band, horizontal visibility $\Delta x=26$ km, optical depth $\tau=0.3$.  We limited our computations to zenith sky brightness even 

 \onecolumn
\newpage
\newpage

\newpage
	
\newpage

\twocolumn
\noindent 
  though our method allows determinations of brightness in other directions.
 This would be useful to predict visibility in large territories of particular astronomical phenomena, like e.g. comets. A complete mapping of the artificial brightness of the sky of a site, like Cinzano \cite{c99b}, using satellite data instead of population data is possible (Cinzano 2000, in prep.). Falchi \& Cinzano \cite{falcin} and Cinzano et al. \cite{cinfal} obtained in 1998 the first maps of the artificial sky brightness from satellite data using a DMSP single-orbit image
and replacing $f$  in eq. (\ref{int1})  with the Treanor Law,  a semi-empirical law which assumes a very simplified model with homogeneous atmosphere, vertical heights small in relation to the horizontal distances, scattering limited to a cone of small angle co-axial with the direct beam and flat Earth.  Differences with our maps mainly arise where Earth curvature play a role in limiting the propagation of light pollution. Our study constitute the natural improvement of their seminal work.

Recommendation 1 of the IAU Commission 50 (Smith 
1979) states that  the increase in sky brightness at 45\degr~ elevation due to
artificial light scattered from clear sky should not exceed 10 per cent of the lowest
natural level in any part of the spectrum between wavelengths 3000{\AA}~ and 
10000{\AA}. So this is the level over which the sky must be considered "polluted". The maps shows that only a few areas in Europe are under the limit of 10 per cent at zenith and some of them could still result in being quite polluted at higher zenith distances.

\section{Conclusions}
\label{s7}

We presented a method to map the artificial sky brightness in large territories in astronomical photometric bands with a resolution of approximately 1 km.  We  computed  the maps with detailed models for the propagation in the atmosphere of the upward light flux measured with DMSP satellites Operational Linescan System. The use of this modelling technique allows us to (i) assess the atmospherical conditions for which the maps are computed giving observable quantities and  (ii) take into account Earth curvature.  This cannot be done properly when using semiempirical propagation laws. The use of satellite data constitutes an improvement over the use of population data to estimate upward flux  because (i) it allows spatially explicit detail of the geographic distribution of emissions and (ii) some polluting sources, like e.g. industrial areas, ports and airports, are not well represented in population data. 

We presented, as an application of  the described method,  the maps of artificial sky brightness in Europe at sea level in V and B bands.
We are extending the maps to the rest of the World in a fore coming World Atlas of Artificial Sky Brightness and preparing predictions for the state of the night sky in future years.

\section*{Acknowledgments}

We are indebted to Roy Garstang of JILA-University of Colorado for his friendly kindness in reading and commenting on this paper, for his helpful suggestions and for interesting discussions.

\appendix

\section{Geometrical relations}
\label{apb}

With the hypotesis of  sea level, geometrical relations from Garstang \cite{g89a}   between quantities in Fig. \ref{geo} became:
 \begin{equation}\label{flat}
\begin{array}{l}
     s=\sqrt{ (u-l)^{2}+4ul \sin^{2}(D/2R_{T})}                                     \\
\mathrm{with}\quad    l=\sqrt{4R_{T}^{2}\sin^{2}(D/2R_{T})}\\          
     h=R_{T} \left(\sqrt{1+\frac{u^{2}+2uR_{T} \cos(z)}{R_{T}^{2}} }-1 \right) \\ 
\omega=\theta+\phi\\     
\mathrm{with}\quad    \theta= \arccos(\frac{q_{1}} {\sqrt{4R_{T}^{2}\sin^{2}(D/2R_{T})} } )\\
\mathrm{}\qquad\quad   \phi=\arccos( \frac{l^{2}+s^{2}-u^{2}}{2ls})\\  
  \psi=\arccos( \frac{q_{2} }{s} )\\ 
\end{array}
\end{equation}
 \begin{displaymath}
\begin{array}{l}
q_{1}= R_{T}(\sin(D/R_{T}) \sin(z) \cos(\beta)+\cos(D/R_{T}) \cos(z))   \\   
q_{2}=u\, \sin(z) \cos(\beta) \sin(D/R_{T}) + q_{3} \\  
q_{3}= u\, \cos(z) \cos(D/R_{T})\! -\! 2R_{T} \sin^{2}(D/2R_{T})\\    
\end{array}
\end{displaymath}
where $R_{T}$ is the Earth curvature radius.

Equations (\ref{xi1}), (\ref{xi2}) have been integrated by Garstang (1989a: eq. (18), (19), (22) and eq. (20), (21)). For sea level and $z<90$\degr~ and $\psi< 90$\degr, they are:
\begin{equation}
\begin{array}{l}
\xi_{1}=\exp \left( - N_{\mathrm{m,0}}  \sigma_{\mathrm{m}}(p_{1}+11.778Kp_{2}) \right)\\
p_{1}=c^{-1}  \sec z \,( 1- \exp(-cu \cos z) 
+ \frac{16 p_{3} \tan^{2} z}{9 \pi 2 c R_{T}} \\
p_{2}=a^{-1}  \sec z \,( 1- \exp(-au \cos z) 
+ \frac{16 p_{4} \tan^{2} z}{9 \pi 2 a R_{T}} \\
p_{3}=(c^{2}u^{2} \cos^{2} z+ 2 cu \cos z + 2) \exp(-cu \cos z)\!-\! 2 \\
p_{4}=(a^{2}u^{2} \cos^{2} z+ 2 au \cos z + 2) \exp(-au \cos z)\!-\! 2 \\
   \end{array}
\end{equation}
\begin{equation}\label{effe}
\begin{array}{l}
\xi_{2}=\exp \left( - N_{\mathrm{m,0}}  \sigma_{\mathrm{m}}(f_{1}+11.778Kf_{2}) \right)\\
f_{1}=c^{-1}  \sec \psi \,( 1- \exp(-cs \cos \psi) 
+ \frac{16 f_{3} \tan^{2} \psi}{9 \pi 2 c R_{T}} \\
f_{2}=a^{-1}  \sec \psi\, ( 1- \exp(-as \cos \psi) 
+ \frac{16 f_{4} \tan^{2} \psi}{9 \pi 2 a R_{T}} \\
f_{3}= (c^{2}s^{2} \cos^{2} \psi+ 2 cs \cos \psi + 2) \exp(-cs \cos \psi)\!-\!2 \\
f_{4}= (a^{2}s^{2} \cos^{2} \psi+ 2 as \cos \psi + 2) \exp(-as \cos \psi)\!-\!2 \\
   \end{array}
\end{equation}
For $\psi=90$\degr~ the reader is referred to the cited paper.

\section{Used DMSP orbits}
\label{tab1}

Orbits of the DMSP satellite F12 used in the composite radiance calibrated image.

  \begin{tabular}{|c|c|c|}
\hline
  199603191827  &		199702061858  &		  199701091752  \\
   199603192009    &	            199702062040       &		  199701091934  \\
   199603192151  &		199702062222  &		   199701092116  \\
   199603211803     &	199702081833  &		 199701111727   \\
   199603211945  &		199702082015  &		 199701111909     \\
   199603212127   &	 199702082157                &	 199701112051  	  \\
  199603231738 &		199702101809  &		  199701131703   \\
  199603231920  &		199702101951  &		 199701131845   \\
  199603232102  &		  199702102133    &	199701132027  	   \\
     199701061828  &          199603171709  &		  199701132209  \\
   199701062010  &	              199603171851   &	199702031752  	   \\
  199701062152     &	199603172215  &		  199702031934  \\
  199701081804  &	             199603181839     &	199702032116  	    \\
   199701081946  &	199603182021  &		 199702032258   \\
  199701082128  &	            199603182203  &		199702051728      \\
  199701101740  &	             199603201815     &	199702051910  	  \\
   199701101922  &	199603201957  &	                     199702052052  	    \\
  199701102104  &	             199603202139     &	199702071703  	  \\
 199701121715    &	199603221750    &	                    199702071845  	    \\
  199701121857  &		199603221932  &	                   199702072027  	  \\
    199701122039&            199603222114                 &	199702072209  	   \\
199701122221  &		199701051841  &	                         199702091821  	  \\
     199702041740 &	199701052023  &	                           199702092003  	    \\
199702041922  &		199701052205  &	                            199702092145  	  \\
199702042104  &		199701071816   &	199701071958	    \\
199702061716    &	199701072140  	&	199701091752  	  \\
 \hline
 \end{tabular}

\label{lastpage}
\bsp
\end{document}